# RECONFIGURABLE MULTITASK AUDIO DYNAMICS PROCESSING SCHEME

*Jun Yang (IEEE Senior Member), Amit S. Chhetri, Carlo Murgia, and Philip Hilmes*

Amazon Lab126, 1100 Enterprise Way, Sunnyvale, CA 94089, USA

**ABSTRACT**

Automatic speech recognition (ASR), audio quality, and loudness are key performance indicators (KPIs) in smart speakers. To improve all these KPIs, audio dynamics processing is a crucial component in related systems. Unfortunately, single-band and existing multiband dynamics processing (MBDP) schemes fail to maximize bass and loudness but even produce unwanted peaks, distortions, and nonlinear echo so that an optimized ASR performance cannot be achieved. It has been a goal in both industry and academia to find a better audio dynamics processing for mitigating these problems. To provide such a desired solution, this paper proposes a novel reconfigurable multitask MBDP scheme through a global optimization framework. Through extensive testing, we show the accuracy and effectiveness of the proposed scheme in terms of bass and loudness maximization, distortion and nonlinear echo reduction, browning-out prevention, and significant ASR performance improvement.

## 1. INTRODUCTION

MBDP plays a very important role not only in audio enhancement systems, but also in acoustic echo cancellation (AEC) systems [1 - 3] of emerging smart speakers [4]. However, due to the poor frequency resolution, fixed center frequencies, fixed bandwidths of filterbank designs and the limited flexibility of adjusting the compression degree in each band, the existing MBDP schemes [5 - 8] cannot precisely reduce peak-to-average ratio (PAR). Consequently, they can neither maximize bass and loudness, nor precisely reduce loudspeaker total-harmonic-distortion (THD), which can produce nonlinear echo and degrade AEC and ASR performance. In addition, the existing MBDP schemes cannot precisely limit audio peaks in each band so that they have browning-out problem for power-limited audio devices.

It has been a goal for long time for the industry and academia to provide an effective MBDP scheme through a global optimization framework, so as to achieve the best compromise between ASR performance, audio quality, loudness, and complexity. However, this well-designed MBDP scheme could not be accomplished from the immediate generalization of all the existing solutions mainly because their related compression degree is not dynamically and flexibly adjustable.

To achieve this goal, this paper proposes and evaluates a reconfigurable volume-dependent MBDP scheme, which includes a reconfigurable filterbank, a flexible multiband compressor (MBC), a multiband limiter (MBL), a mixer, as well as a full-band limiter (FBL). The major feature of the proposed scheme is a global optimization processing by jointly maximizing MBDP, AEC, and ASR performance over a large-scale database. The given theoretical analyses, and subjective and objective test results show that the proposed scheme can not only maximize bass and loudness, and deliver good audio performance, but it can also reduce nonlinear echo so as to offer a significant improvement for AEC and ASR performance in emerging smart speakers.

The rest of this paper is organized as follows. Section 2 mainly presents the proposed system and reconfigurable filterbank algorithm. The emphasis of Section 3 is to present the proposed flexible MBC algorithm. By using various test results, Section 4 mainly shows that smart speakers implemented with the proposed MBDP scheme can significantly reduce distortion, maximize bass and loudness, and significantly reduce false-rejection-rate (FRR) which measures the percentage of the missed ASR keyword (aka wakeword, or wake-up word) commands. Section 5 will provide conclusions and further discussions.

## 2. THE PROPOSED SYSTEM AND RECONFIGURABLE FILTERBANK ALGORITHM

Figure 1 illustrates the proposed reconfigurable volume-dependent MBDP scheme, which is integrated with an AEC system. The HPF, RES, NR, and AGC denote acronyms for high-pass filter, residual echo suppression, noise reduction, and automatic gain control, respectively. Their algorithmic details can be found in our previous works [2, 3, 9].

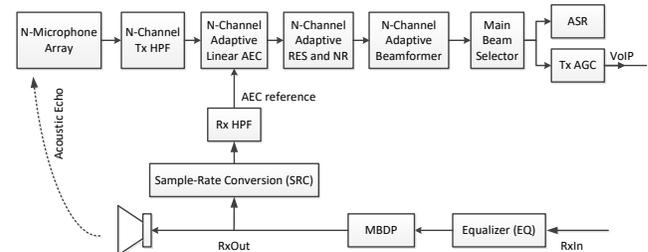

Figure 1: The proposed MBDP scheme integrated with AEC system.

Figure 2 illustrates the detailed processing stages of the proposed reconfigurable volume-dependent MBDP scheme.

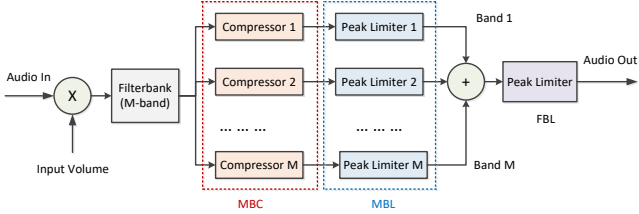

Figure 2: The details of the proposed MBDP scheme.

In Figure 2, the "Input Volume" block can either boost or attenuate the raw audio. Before presenting the details of "Filterbank" and MBC algorithms, we first address the MBL and FBL blocks. More specifically, the M-band MBL algorithm consists of *M* "Peak Limiter" blocks. Each "Peak Limiter" [9] uses the same algorithm to measure peak and control that no peak will exceed the predefined corresponding peak threshold. This feature can precisely limit the large peaks across bands so as to prevent browning-out problem for power-limited audio devices.

In Figure 2, the *M*-band output signals of MBL component are combined into a full-band signal. An additional FBL is employed to guarantee that the combined full-band audio does not create unwanted peak levels. More importantly, the peak threshold of FBL can be adjusted to make the loudspeaker work in its linear dynamic range so as to reduce nonlinear echo and improve AEC and ASR performance.

In Figure 2, the "Filterbank" block splits full-band audio into M bands (e.g., M = 2, 3, …). As a further illustration, a reconfigurable filterbank is presented in Figure 3, which can reconstruct the signal perfectly. The center frequencies, bandwidths, and number of bands are adjustable in real time.

For the sake of simplicity, only 2-band, 3-band, and 4-band architectures are shown in Figure 3. As a matter of fact, the architecture for M-band (M > 4) is a scalable version of 4-band in Figure 3.

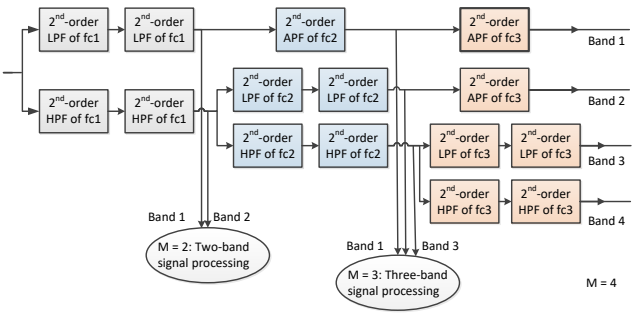

Figure 3: The proposed reconfigurable filterbank.

In Figure 3, $fc1$, $fc2$, $fc3$, … are tunable crossover frequencies with the requirement of 0.0 Hz < $fc1$ < $fc2$ < $fc3$ < … < $fs/2$, where $fs$ is sampling rate (e.g. 48 kHz), so that the bandwidths and center frequencies are adjustable. The LPF and APF stand for low-pass filter and all-pass filter, respectively. APF is designed to match phase response between bands. Two cascaded LPFs, two cascaded HPFs, and one APF have the same phase response if they have the same crossover frequency. Therefore, the outputs of each band are in-phase.

The values of crossover frequencies are related to the loudspeaker's characteristics. Figure 4 shows magnitude responses of a four-band filterbank by using the example of $fc1$ = 70 Hz, $fc2$ = 375 Hz, and $fc3$ = 3750 Hz. Obviously, all the stop-bands of magnitude responses have no ripples, which results in a high signal-to-noise ratio. Moreover, the proposed four-band filterbank is perfectly reconstructible according to the magnitude response of "Sum of 4 Bands".

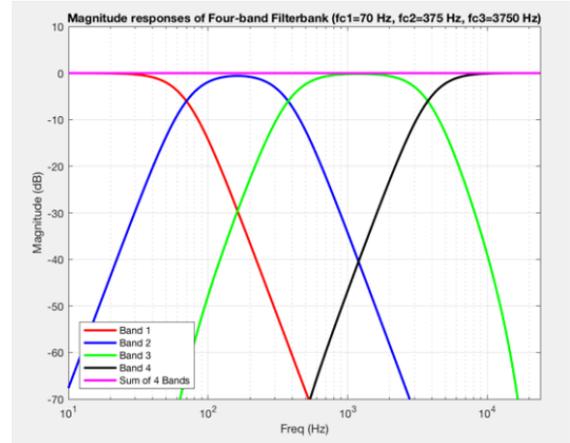

Figure 4: Magnitude responses of the proposed 4-band filterbank.

## 3. THE PROPOSED FLEXIBLE MBC ALGORITHM

The proposed M-band MBC algorithm in Figure 2 consists of *M* Compressor blocks. All Compressors use the same processing stages as described in Figure 5.

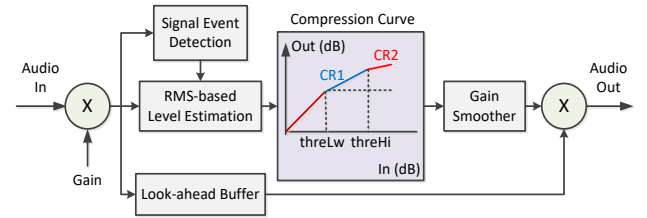

Figure 5: The proposed flexible compressor algorithm.

The "Signal Event Detection" and "RMS-based Level Estimation" blocks in Figure 5 are described in [9]. The "Compression Curve" is defined by four parameters, namely, low threshold (i.e., *threLw* dB), high threshold (i.e., *threHi* dB), and compression ratios *CR1* and *CR2* with the requirements of *threLw* < *threHi* ≤ 0.0 dB and 1.0 ≤ *CR1* ≤ *CR2*. The compression ratio, *CR*, is defined as a ratio between the input level difference and the output level difference as shown in the following.

$$CR = \frac{In1\,(dB) - In2(dB)}{Out1(dB) - Out2(dB)} \quad (1)$$

The four parameters can be adjustable flexibly and independently among bands and audio volumes in real-time. With the help of reconfigurable filterbank, the proposed flexible MBC algorithm can precisely reduce band level only when audio dynamics in the band exceed the corresponding threshold, and can precisely reduce PAR over frequency bands so as to maximize bass and loudness and precisely reduce loudspeaker THD.

An intermediate adaptive gain $p(k, n)$ in the $n$-th frame and $k$-th band can be calculated as follows by using input level $L(k, n)$ (in dB) and "Compression Curve".

If $L(k, n) <$ threLw, then it is a linear processing.
$$p(k, n) = 0.0 \text{ dB} \quad (2)$$
If threLw $\leq L(k, n) <$ threHi, then
$$p(k,n) = \left(1 - \frac{1}{CR1}\right) \times \left(threLw - L(k,n)\right) \quad (3)$$
If $L(k, n) \geq$ threHi, then
$$p(k,n) = \left(1 - \frac{1}{CR2}\right) \times \left(threHi - L(k,n)\right) \quad (4)$$

The linear adaptive gain $q(k, n)$ can be obtained as follows.
$$q(k,n) = 10^{p(k,n)/20.0} \quad (5)$$

The "Gain Smoother" block in Figure 5 is used to reduce the variation of the adaptive gain. Multiplying the delayed block of input signals (i.e., the audio samples in the look-ahead buffer) by the smoothed linear gain in a way of sample-by-sample results in a well and smoothly controlled output level.

## 4. EVALUATION RESULTS

This section presents evaluation results of the proposed reconfigurable multitask MBDP scheme in terms of distortion reduction, bass and loudness maximization, as well as the improvement of AEC and ASR performance due to nonlinear echo reduction.

### 4.1. Audio distortion reduction

Figure 6 shows waveform (top), i.e., amplitude (dB) versus time, and spectrogram (bottom), i.e., frequency (Hz) versus time, of a test input signal. Figure 7 shows waveform (top) and spectrogram (bottom) of output signal of a traditional MBDP scheme. Figure 8 shows waveform (top) and spectrum (bottom) of the output signal of the proposed MBDP scheme. Figures 6, 7, and 8 use the same plot-scales.

It can be seen from Figure 7 that the traditional scheme produces the unwanted peaks and overshoot distortions, which can degrade sound quality, could shut down some battery-driven audio devices, and produce nonlinear echo. In comparison, Figure 8 demonstrates that the proposed reconfigurable MBDP scheme has removed these unwanted peaks and overshoot distortions by the proposed MBL and FBL algorithms. The successful removal of these unwanted peaks and distortions can not only improve the sound quality but also can help precisely limit the peaks to prevent browning-out problem, avoid clipping distortion, and reduce nonlinear echo. In addition to these acoustical tests, our extensive listening test results have also shown that the audio signal processed by the proposed MBDP scheme also sounds much better than that processed by traditional MBDP scheme in terms of distortion and bass.

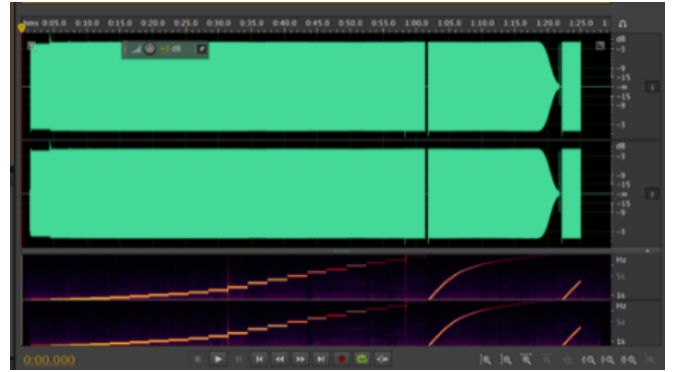
Figure 6: Test input signal.

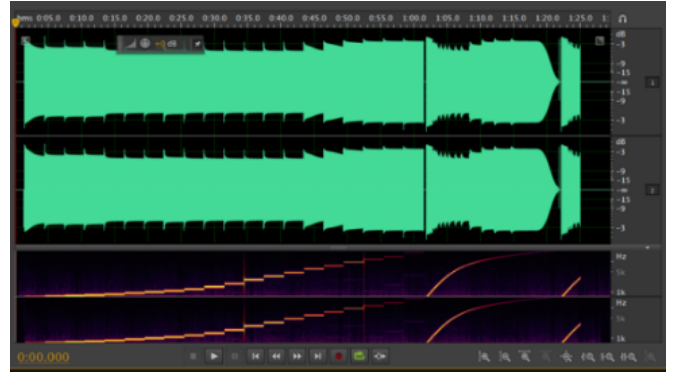
Figure 7: Output signal of a traditional MBDP scheme.

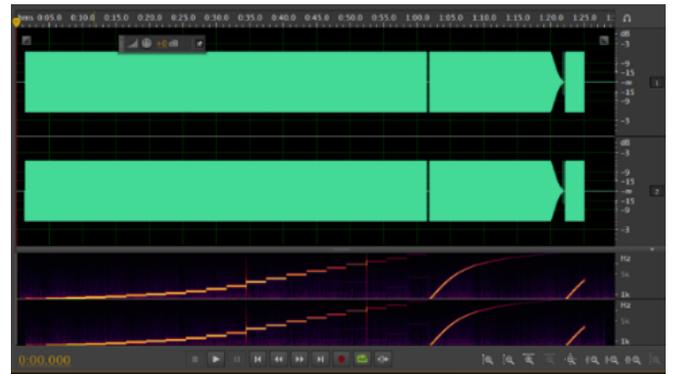
Figure 8: Output signal of the proposed MBDP scheme.

### 4.2. Bass and loudness enhancement

As given in Sections 2 and 3, the proposed reconfigurable MBDP scheme can precisely reduce the PAR over frequency bands by the novel reconfigurable filterbank and flexible MBC algorithms. This band-accurate PAR reduction is a key factor to maximize the bass and loudness as shown in the following test results.

Figure 9 shows long-term power spectral density (PSD) plots of traditional (black plot) and the proposed (blue plot)

MBDP schemes. Both schemes use the same test input signal. It can be seen that the proposed reconfigurable MBDP scheme increases the bass by approximately 5 dB and increases loudness up to 1.8 kHz. Extensive real-time listening tests have demonstrated that audio processed by the proposed reconfigurable MBDP scheme sounds much better and louder than that processed by the traditional MBDP scheme.

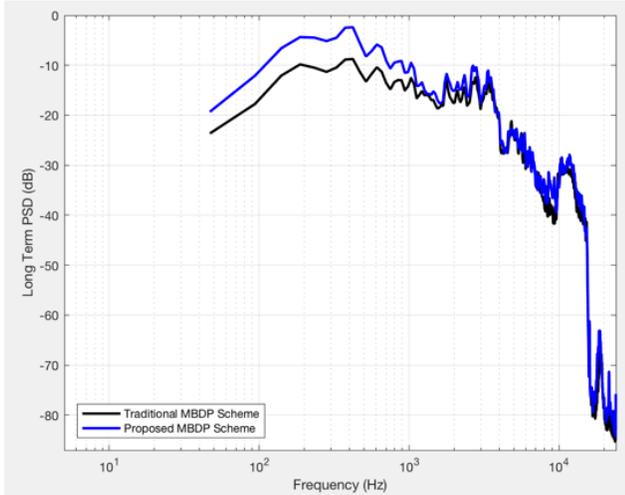

Figure 9: Long-term PSD plots of traditional (black) and the proposed (blue) MBDP schemes.

### 4.3. Wake-up word recognition improvement

Figure 10 shows average THD ratio versus frequency measured over a prototype audio device for 5 types of audio volumes; the larger the audio volume, the higher the loudspeaker THD. Obviously, loudspeaker THD changes over both frequency and audio volume. Loudspeaker THD can produce nonlinear echo and significantly reduce AEC and ASR performance.

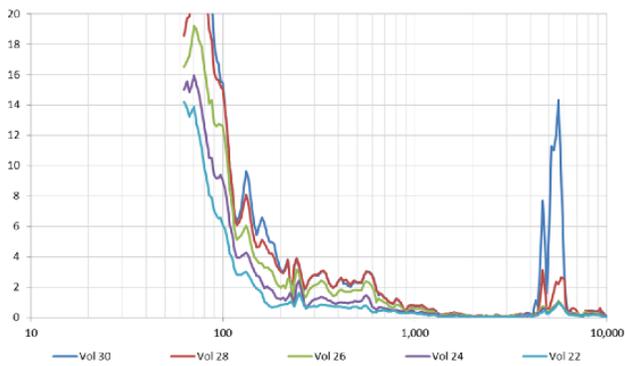

Figure 10: Average THD ratio (%) versus frequency (Hz).

One major factor of ASR performance degradation is nonlinear echo. The proposed reconfigurable MBDP scheme can reduce nonlinear echo in three ways as pointed out in previous sections: (1). the reconfigurable filterbank and flexible MBC algorithms can precisely reduce the PAR over frequency bands and audio volumes so as to precisely reduce loudspeaker THD, (2). the proposed MBL and FBL can make loudspeaker work in its linear dynamic range, (3). the proposed FBL can prevent output audio from clipping so as to minimize distortion of digital signal. This suggests that a significant improvement of AEC and ASR performance can be achieved by our proposed MBDP scheme as shown in the following test results.

Figure 11 shows FRR versus playback volume for traditional (black plot) and proposed (blue plot) MBDP schemes on this device. It is evident that the proposed MBDP scheme significantly reduces FRR and outperforms traditional MBDP scheme.

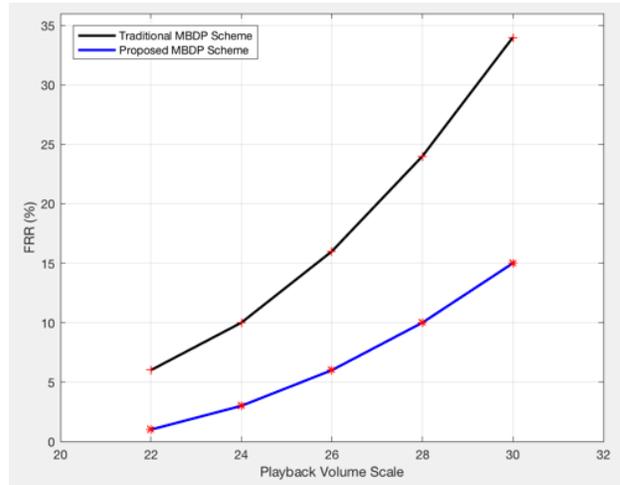

Figure 11: FRR (%) versus playback volume.

### 6. CONCLUSONS

The major advantages of each key processing block of the proposed processing scheme can be summarized as follows. The proposed reconfigurable filterbank algorithm is perfectly reconstructible and provides user with adjustable bandwidths, center frequencies, and number of bands. With the help of reconfigurable filterbank algorithm, the proposed flexible MBC scheme can precisely reduce PAR in each band separately so as to maximize bass and loudness, and precisely reduce loudspeaker THD as well as nonlinear echo. The proposed MBL scheme can guarantee audio peaks of each band are below the predefined peak thresholds so as to control the power limited loudspeakers and prevent them from browning-out. Correspondingly, the FBL prevents audio from overflow, clipping, and distortion and makes loudspeaker work in its linear dynamic range so as to deliver good audio and reduce nonlinear echo.

By addressing various types of audio dynamics, settings, and conditions, theoretical analyses, subjective and objective test results conducted on a real-time hardware computing platform have shown that the proposed reconfigurable multitask MBDP scheme can offer a significant improvement for bass, loudness, audio quality, AEC, and ASR performance in emerging smart speakers.

# 6. REFERENCES

[1] Jason Wung, "A System Approach to Multi-Channel Acoustic Echo Cancellation and Residual Echo Suppression for Robust Hands-free Teleconferencing," Ph.D. Dissertation, School of Electrical and Computer Engineering, Georgia Institute of Technology, May 2015

[2] Jun Yang and Philip Hilmes, "Dynamics and Periodicity Based Multirate Fast Transient-Sound Detection," the 26th European Signal Processing Conference (EUSIPCO), Rome, Italy, pp. 2463 - 2467, Sept. 3-7, 2018

[3] Jun Yang, "Multilayer Adaptation Based Complex Echo Cancellation and Voice Enhancement," ICASSP 2018, Calgary, Albert, Canada, pp. 2131 - 2135, April 15-20, 2018

[4] Amit Chhetri, Philip Hilmes, Trausti Kristjansson, Wai Chu, Mohamed Mansour, Xiaoxue Li, and Xianxian Zhang, "Multichannel Audio Front-End for Far-Field Automatic Speech Recognition", 26th European Signal Processing Conference (EUSIPCO), Rome, Italy, pp. 1541 - 1545, Sept. 3-7, 2018

[5] Will Pirkle, "Chapter 13 Dynamics Processing" of "Designing Audio Effect Plug-Ins in C++: With Digital Audio Signal Processing Theory," First Edition; ISBN-13: 978-0240825151; ISBN-10: 0240825152; Oct 15, 2012

[6] Udo Zolzer, "Chapter 7 Dynamic Range Control" in "Digital Audio Signal Processing," Second Edition, Wiley; ISBN-13: 978-0470997857; ISBN-10: 0470997850; July 28, 2008.

[7] Dimitrios Giannoulis, Michael Massberg, and Joshua D. Reiss, "Digital Dynamic Range Compressor Design – A Tutorial and Analysis," *J. Audio Eng. Soc.*, Vol. 60, No. 6, June 2012, PP. 399 - 408

[8] Jicai Liang, Song Gao, Yi Li, "Research on Dynamic Range Control used to Audio Directional System," *2011 Internal Conference on Mechatronic Science, Electric Engineering and Computer*, August 19-22, 2011, PP. 498 - 501, Jilin, China

[9] Jun Yang, Philip Hilmes, Brian Adair, and David W. Krueger, "Deep Learning Based Automatic Volume Control and Limiter System," ICASSP 2017, New Orleans, USA, pp. 2177 - 2181, March 5 - 9, 2017